\documentclass[aps,twocolumn]{revtex4} 

\usepackage{epsf}

\begin{document}

\title{Density Matrix Renormalisation Group Calculations for Two-Dimensional 
 Lattices: An Application to the Spin-Half and Spin-One Square-Lattice 
 Heisenberg Models}

\author{D.J.J. Farnell} 

\address{Imaging Science and Biomedical Engineering (ISBE), 
 University of Manchester, Oxford Road, Manchester M13 9PT, United Kingdom}

\date{\today}

\begin{abstract}
A new density matrix renormalisation group (DMRG) approach is 
presented for quantum systems of two spatial dimensions. In 
particular, it is shown that it is possible to create a 
multi-chain-type 2D DMRG approach which utilises previously 
determined system and environment blocks {\it at all points}.
One firstly builds up effective quasi-1D system and environment 
blocks of width $L$ and these quasi-1D blocks are then used to 
as the initial building-blocks of a new 2D infinite-lattice 
algorithm. This algorithm is found to be competitive with 
those results of previous 2D DMRG algorithms and also of 
the best of other approximate methods. An illustration 
of this is given for the spin-half and spin-one Heisenberg 
models on the square lattice. The best results for the 
ground-state energies per bond of the spin-half and 
spin-one square-lattice Heisenberg antiferromagnets
for the $N = 20 \times 20$ lattice using this 
treatment are given by $E_g/N_B = -0.3321$ and 
$E_g/N_B = -1.1525$, respectively. 

\begin{flushleft}
\end{flushleft}
\end{abstract}

\maketitle


\section{Introduction}

The subject of one-dimensional lattice quantum spin systems at 
zero-temperature contains a large number of exact solutions, for 
example, via the Bethe Ansatz solution \cite{ba1,ba2,ba3,ba4}. 
However, such exact solutions have not as yet been universally 
obtained for systems of quantum spin number, $s \ge 1/2$, or
for lattices of larger than one spatial dimension, or indeed 
for frustrated spin systems. It is noted however that exact solutions for 
such systems do exist for various specialised cases. Recent density 
matrix renormalisation group (DMRG) calculations \cite{dmrg1} have 
been extremely useful in determining the ground- and excited-state 
properties of a whole host of one-dimensional (1D) or quasi-1D spin 
systems to extremely high accuracy. The DMRG method is not 
limited by the magnitude of the quantum spin number or by the 
presence of frustration, and DMRG has been used to determine 
the non-zero temperature and dynamical properties of many 
quantum systems. 

It is perhaps still fair to say that the full power of the DMRG 
method has largely been restricted to one-dimensional systems, 
although recent calculations have very successfully extended 
\cite{dmrg_2D_1,dmrg_2D_2,dmrg_2D_3,dmrg_2D_999,dmrg_2D_4,dmrg_2D_5,dmrg_2D_6} 
previous DMRG calculations for one-dimensional quantum systems to 
two spatial dimensions (2D). Perhaps the simplest such 2D 
DMRG approach is the ``multi-chain'' approach
\cite{dmrg_2D_1,dmrg_2D_2,dmrg_2D_3,dmrg_2D_5} in which, during 
the infinite-lattice algorithm, ones keeps the width of the 
lattice constant and then one grows the lattice height by
adding whole rows or partial rows of sites. The multi-chain 
finite-lattice algorithm may then be implemented and one 
repeatedly sweeps through all lattice sites until a fixed point 
of the renormalisation group (RG) 
process is reached. This is in direct analogy with
the traditional one-dimensional algorithm, and so (in some sense)
one has mapped the 2D problem onto an effective 1D problem.
This multi-chain approach has been seen to work well 
for relatively small-sized lattices in Ref. \cite{dmrg_2D_5} 
in which whole lines of sites are added at a time. Indeed, 
a ``finite-lattice sweep'' was performed after each addition
of a line of sites (and not just after the final $N = L \times 
L$ lattice had been created) in this calculation 
\cite{dmrg_2D_5}, and this was seen to greatly enhance the 
accuracy of the method.

Another approach \cite{dmrg_2D_4} was to split the two-dimensional
lattice into four pieces, namely, two-dimensional system 
and environment blocks separated by a lines of spins which
itself was split into distinct blocks, namely, one-dimensional 
system and environment blocks, and free spin(s). This ``four block''
approach is a truly two-dimensional algorithm in the sense
that one is able to grow the two-dimensional lattice in a 
controlled manner using the results (i.e., 1D and 2D blocks)
of previous iterations and a lattice of arbitrary size may 
be thus finally obtained. However, this approach suffers from 
a number of problems such as, for example, the need to 
diagonalise very large density matrices for the two-dimensional 
blocks for relatively modest numbers of states. We also note 
that these density matrices for the 2D blocks may also be 
highly singular. 

Another recent and very successful 2D DMRG \cite{dmrg_2D_5} 
approach used two blocks at opposite {\it corners} of a square 
lattice with two ``free'' spins placed at particular positions
along the boundary between the system and environment
blocks. It was seen for this method that there was no 
need to set the width of the lattice to be constant and
that previously defined blocks for a given $N=L \times L$
lattice could be used in the following DMRG iterations
to form lattices of size $N'=L+1 \times L+1$. Furthermore, 
this approach was found to be very efficient
for the (relatively small) lattices ($N=L \times L$ with
$L \le 12$) considered, and excellent results for
the ground-state energy per bond of the spin-half
Heisenberg model on the square and triangular lattices
were obtained. An application of the DMRG method
for highly anisotropic two-dimensional systems was
also recently performed \cite{dmrg_2D_6}. 

In this article a new 2D algorithm is presented which 
builds up two-dimensional lattices in a fast, simple 
and efficient manner. Quasi-1D blocks of up to a given length, 
$L$, are determined and thereafter the width is kept 
constant -- in analogy with previous multi-chain
approaches. However, these quasi-1D blocks are then used
to form the first initial steps of an algorithm which uses 
2D blocks {\it of various shape} and which have been determined 
during earlier DMRG steps in order to finally build up a 
lattice of size $N=L\times L$. Lattices of arbitrary size
may thus be considered, although this `size' is pre-set at 
the start of the DMRG calculation. This is in marked 
contrast to previous multi-chain approaches. 
The 2D ``finite-lattice'' DMRG algorithm has also 
been implemented here in order to arrive at a
fixed point of the renormalisation group process, and
this part of the algorithm is similar (if not identical)
to previous multi-chain algorithms.  

Note that the details of the DMRG approach are
deferred until the appendix. However, it is 
important to note that the current approach does not 
utilise a row 
of sites or partial row in order to build up
the lattice in the infinite-lattice algorithm.
Rather, it does use previously defined blocks for
the environment block which have both width
and height at all points. The total system size 
is thus incremented by more than a single line 
of sites at a time, although the width is kept 
constant.

\section{The Heisenberg model for the square lattice}

The Heisenberg model is given by 
\begin{equation}
H = \sum_{\langle i,j\rangle} {\bf s}_i ~ \cdot ~ {\bf s}_j
~~ ,
\label{eqn1}
\end{equation}
and where ${\langle i,j\rangle}$ runs over each nearest-neighbour
bond on the square lattice (counting each bond once and once 
only) and open boundary conditions are assumed. The results of the 
best of other approximate methods for the spin-half case
\cite{qmc4,series1,swt,ccm1} predict that $E_g/{N_B} 
\approx -0.3347$ (where $N_B$ is the number of bonds) 
and that 61-62\% of the classical ordering remains 
for the quantum system in the infinite-lattice limit, 
$N \rightarrow \infty$. The results of the 
best of other approximate methods for the spin-one case
\cite{series1,ccm2} predict that $E_g/{N_B} 
\approx -1.164$ (where $N_B$ is the number of bonds) 
and that 80-81\% of the classical ordering remains 
for the quantum system in the infinite-lattice limit.

\section{Results}

\begin{figure}
\epsfxsize=7cm
\centerline{\epsffile{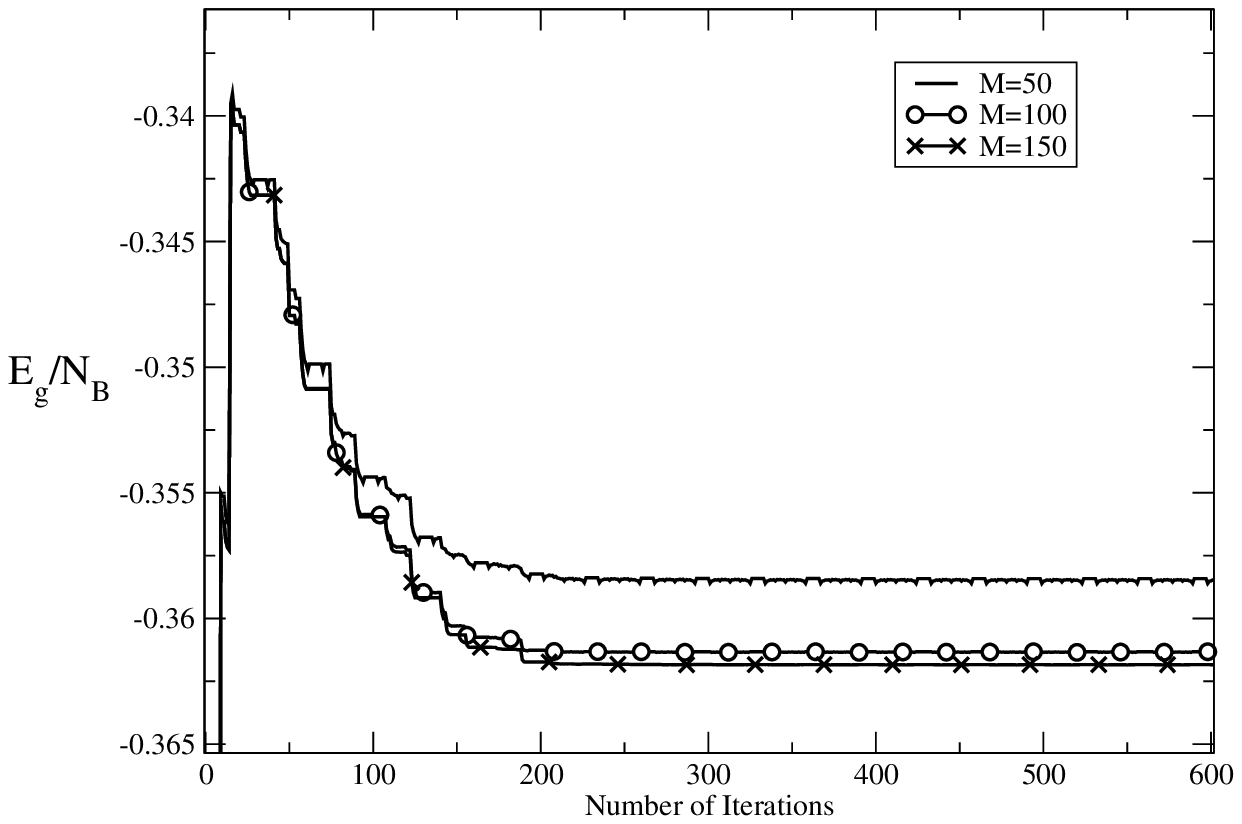}}
\vspace{-2cm}
\caption{The DMRG ground-state energies per bond of
the spin-half Heisenberg on a $N=6\times6$ square
lattice for $M=\{50,100,150\}$. Note that each 
``iteration'' indicates a 2D DMRG ``move'' described 
in detail in the appendix. Note that the infinite-lattice
algorithm ends after 22 iterations (close to the peak in
the $E_g/N_B$) and the results thereafter refer to
those for the finite-lattice algorithm.}
\label{fig7}
\end{figure}

A plot of the ground-state energy per bond for the 
$s=1/2$ Heisenberg antiferromagnet for the 
$N=6\times6$ lattice for $M=\{50,100,150\}$
is presented in Fig. \ref{fig7}. 
It is seen that the trend of the ground-state energy 
is to decrease with the number of iterations and 
the minimal solution is quickly reached. Note
however that edge effects are seen where, for example,
the ground-state energy is slightly lower for
moves 1-3 of the algorithm presented above. 
Also, the infinite-lattice algorithm ends after
22 ``moves'' indicated on the abscissa
as ``iterations'' in Fig. \ref{fig7} 
and we note that the ground-state 
energy per bond has a peak at this 
point. Thus, the finite-lattice 
algorithm reduces the ground-state
energy per bond by about 6\% to 8\%.
Slightly larger gains (up to about 10\%) 
from the finite-lattice algorithm are 
possible for the larger lattices. 

\begin{figure}
\epsfxsize=7cm
\vspace{-2.7cm} 
\centerline{\epsffile{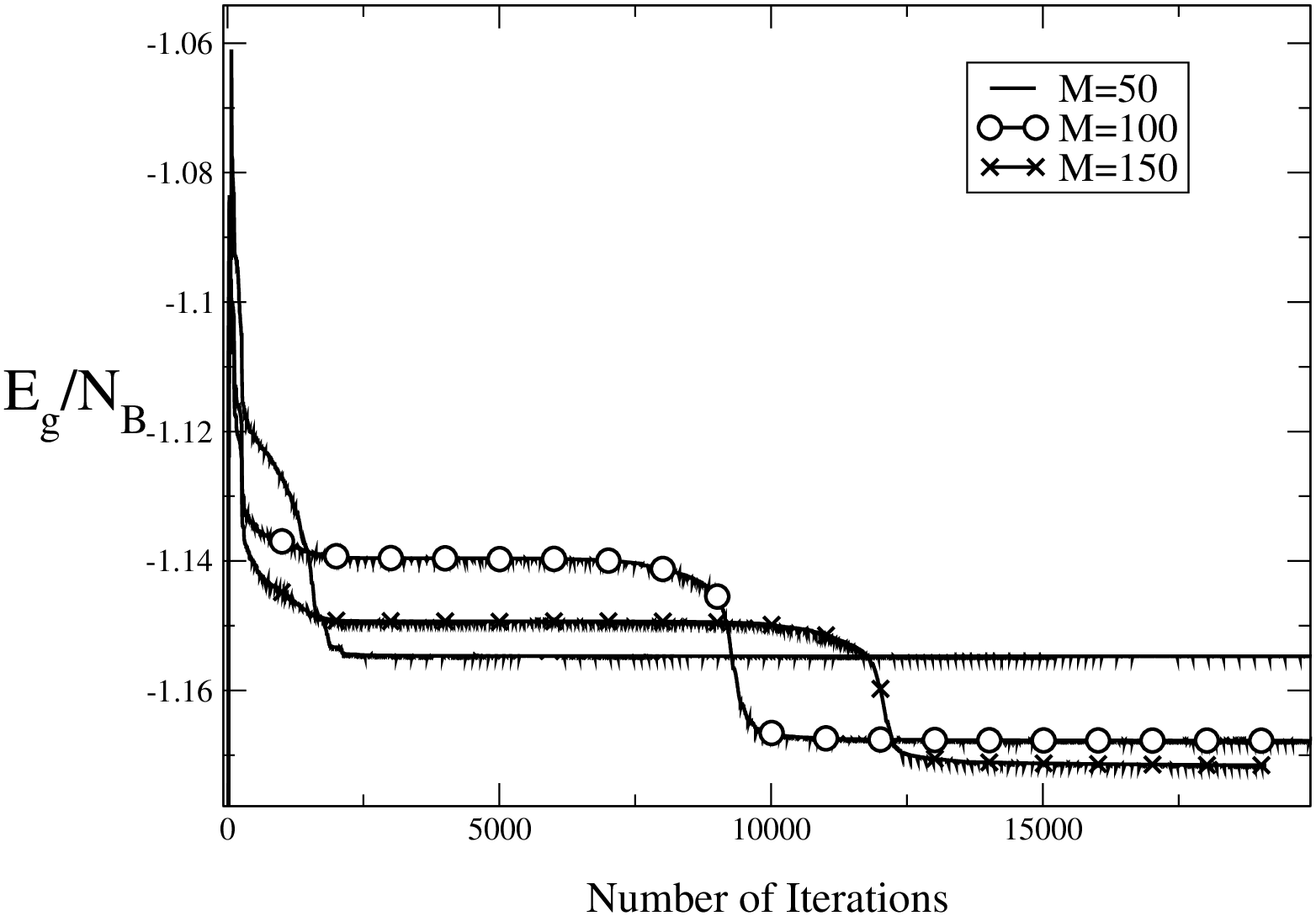}}
\caption{The DMRG ground-state energies per bond of
the spin-one Heisenberg on a $N=12\times12$ square
lattice for $M=\{50,100,150\}$. Note that each 
``iteration'' indicates a 2D DMRG ``move'' described 
in detail in the appendix. Note that the infinite-lattice
algorithm ends at the peak in $E_g/N_B$ and the results 
thereafter refer to those for the finite-lattice 
algorithm.}
\label{figNEW}
\end{figure}

A value for the ground-state energy is taken 
from one of the sites near to the centre of the 
2D square lattice for all of the $N=L \times L$
lattices considered here. By contrast, it is noted 
that the number of iterations needed to obtain
convergence increases strongly with increasing $N$. 
It is observed that the number of iterations 
necessary can be as large as 20000 or more 
for $L=12$ and $L=20$. Indeed, it was seen 
that the ground-state energy sometimes decreases 
in a step-like fashion (even 
after 10000 or more iterations large step-like 
changes were sometimes observed). An example of this is given 
in a plot of the ground-state energy per bond for the 
$s=1$ Heisenberg antiferromagnet for the  
$N=12\times12$ lattice for $M=\{50,100,150\}$
of Fig. \ref{figNEW}. It would be interesting 
to test whether such step-wise changes occur
for other 2D DMRG algorithms and if this were
a characteristic of implementing the DMRG 
method in two spatial dimensions. Finally, an
explicitly assumption of these simulations
was that they had been run long enough in 
order to reach the RG fixed-point. 

Note that the infinite-lattice algorithm ends 
after about 80 ``moves'' in Fig. \ref{figNEW} 
(which are again indicated on the abscissa as 
``iterations'') and we note that the ground-state 
energy per bond has a peak at this 
point. Thus, the finite-lattice 
algorithm reduces the ground-state
energy per bond by about 8\% to 10\%.

These results may also be compared to results of 
previous DMRG calculations \cite{dmrg_2D_5} in 
Tab. \ref{tab1} for the spin-half Heisenberg 
model on square lattices of varying size. 
Two algorithms were used in Ref. \cite{dmrg_2D_5},
which were, namely, a 2D algorithm in which the 2D system 
and environment blocks are grown from the corners and a 
multi-chain approach (also described in Refs. 
\cite{dmrg_2D_1,dmrg_2D_2,dmrg_2D_3}) in which
whole lines of sites were added at a time during the
infinite-lattice approach. 
The results of these approaches shall henceforth 
be referred in this article to as DMRG (a) and DMRG 
(b) results, respectively, and the results of the present 
approach using the infinite-lattice algorithm
shall be referred to as DMRG (c) results.

It is seen from Tab. \ref{tab1} that we obtain reasonable 
correspondence between the present DMRG (c) results and the 
previous DMRG (a) and (b) results for all  the lattices considered. 
A simple `heuristic' extrapolation of these results in the limit
$M \rightarrow \infty$ by assuming that the ground-state
energy scales as a ``power-law'' with respect to $M$,
which ought to be the case for a gapless system. The best
fit of such a power-law dependence was determined and the
extrapolated values were determined. Note that this
simple treatment gave an extrapolated value for the
ground-state energy per bond of $E_g/N_B = -0.36219$
for the $N=6 \times 6$ lattice. This result is in good 
agreement with the extrapolated value for the ground-state
energy of the spin-half square-lattice Heisenberg 
antiferromagnet of Ref. \cite{dmrg_2D_5}, namely, $E_g/N_B = -0.36212$.
The extrapolated result for the ground-state energy per 
bond of the spin-half antiferromagnet on the 
$N=20 \times 20$ lattice is given by $E_g/N_B = -0.3321$.
This result for the $N=20 \times 20$  lattice is in 
reasonable agreement with the results of the best of 
other approximate methods \cite{qmc4,series1,swt,ccm1}. 
These methods predict that $E_g/{N_B} \approx -0.3347$ 
in the infinite-lattice limit, $N \rightarrow \infty$,
and note that this result ought to be quite close
to the result of the $N=20 \times 20$ lattice, even 
with closed boundary conditions. (Note that no 
extrapolations of the 2D DMRG results in the limit 
$N \rightarrow \infty$ were determined due to the 
small number of data points and the fact that a 
relatively large lattice was considered anyway, 
although such an extrapolation is possible to 
perform.) 

\begin{table}
\caption{DMRG results for the ground-state energy per bond 
of the spin-half Heisenberg antiferromagnet on the square lattice for 
lattices of size $N = L \times L$ with $L= \{6,12,20\}$
and for number of DMRG states $M= \{50,100,150\}$. 
The new results are referred to as DMRG (c) results and these
results are compared to two previous DMRG two-dimensional approaches 
(see Ref. \cite{dmrg_2D_5}) 
referred to as DMRG (a) and DMRG (b) results, respectively. It is seen 
that the present DMRG results are in reasonable agreement with these previous
2D DMRG results -- especially for the $N=12\times12$ lattice.}
\begin{tabular}{|l|c|c|c|c|} \hline
                    &$N$          &$M=50$  
                    &$M=100$       &$M=150$          \\ \hline\hline
DMRG (a)            &$6\times6$   &$-$0.361972          
                    &$-$0.362096  &--               \\ \hline
DMRG (b)            &$6\times6$   &$-$0.361919          
                    &$-$0.362089  &--               \\ \hline
DMRG (c)            &$6\times6$   &$-$0.35847              
                    &$-$0.36135   &$-$0.36184 \\ \hline\hline

DMRG (a)            &$12\times12$ &$-$0.337374            
                    &$-$0.341588  &--               \\ \hline
DMRG (b)            &$12\times12$ &$-$0.332574             
                    &$-$0.338833  &--              \\ \hline
DMRG (c)            &$12\times12$ &$-$0.33260              
                    &$-$0.33757   &$-$0.33984       \\ \hline\hline
DMRG (c)            &$20\times20$ &$-$0.31763              
                    &$-$0.32441   &$-$0.32679               \\ \hline
\end{tabular}
\label{tab1}
\end{table}

Results for the spin-one Heisenberg model are 
also shown in Tab. \ref{tab3} for square lattices of
size $N=12 \times 12$ and $N=20 \times 20$. These
these results were again extrapolated in the limit
$M \rightarrow \infty$ and we obtained an extrapolated
result of $E_g/N_B = -1.1525$ for the spin-one 
antiferromagnet on the $N=20 \times 20$ lattice.
This result for $N=20 \times 20$ the lattice is
seen to be in modest agreement with those results 
of the best of other approximate methods \cite{series1,ccm2} 
for both the ground-state energy per bond, evaluated in 
the limit $N \rightarrow \infty$, namely, $E_g/{N_B} 
\approx -1.164$.  

It may also be seen from Tab. \ref{tab1} that the DMRG (a) results
of the 2D algorithm of Ref. \cite{dmrg_2D_5} appear to
perform better in terms of the ground-state energy per
bond of the spin-half Heisenberg model than either the 
DMRG (b) and the present DMRG (c) approaches for all of 
the lattices considered. It is thus interesting to consider
why different DMRG approaches for exactly the same
model and lattice at equivalent levels of approximation
yield differing results.  It is known \cite{global1} from the simulation 
of one dimensional chains with periodic boundary conditions 
that the ground state does not become fully translationally 
invariant, thus indicating that the ``global minimum" 
obtained may be somewhat dependent on the build up considered. 
However, even for the same topological buildup 
and the same final number of states kept, the outcome may 
depend qualitatively on which way the number of states 
kept was increased during the sweeping process of the 
finite-lattice algorithm \cite{global2}.

\begin{table}
\caption{New DMRG results for the ground-state energy 
per bond of the spin-one Heisenberg antiferromagnet on the 
square lattice for lattices of size $N = L \times L$ with 
$L= \{12,20\}$ and for number of DMRG states $M= \{50,100,150\}$. 
The DMRG results are compared to those 
results of SWT \cite{series1}, cumulant series expansions \cite{series1}, 
and the CCM \cite{ccm2} for the infinite-lattice limit.}
\begin{center}
\begin{tabular}{|l|c|c|}  \hline 
                    &$N$          &$E_g/N_B$    \\ \hline\hline
DMRG(c) $M=50$  
                    &$12\times12$ &$-$1.15472   \\ \hline
DMRG(c) $M=100$  
                    &$12\times12$ &$-$1.16780   \\ \hline
DMRG(c) $M=150$  
                    &$12\times12$ &$-$1.1715    \\ \hline
DMRG(c) $M=50$ 
                    &$20\times20$ &$-$1.10654   \\ \hline
DMRG(c) $M=100$ 
                    &$20\times20$ &$-$1.13939   \\ \hline
DMRG(c) $M=150$  
                    &$20\times20$ &$-$1.1462     \\ \hline
CCM                 &$\infty$     &$-$1.1646    \\ \hline
SWT
                    &$\infty$     &$-$1.1641   \\ \hline
Series Expansions
                    &$\infty$     &$-$1.16395(1)\\ \hline
\end{tabular}
\end{center}
\label{tab3}
\end{table}

\section{Conclusions}

In this article a new density matrix renormalisation group (DMRG) 
algorithm was presented which determines the properties of 
two-dimensional square-lattice spin-half and spin-one Heisenberg 
antiferromagnets. This algorithm builds up 2D lattices of 
arbitrary size in a straightforward manner, and, although 
conceptually similar to earlier multi-chain approaches for 
2D lattices, the method differs from these earlier approaches in 
practice because one firstly builds up effective quasi-1D system 
and environment blocks of width $L$. These quasi-1D blocks 
were then used to form the initial steps of a infinite-lattice 
algorithm in order to build a lattice of arbitrary, although 
(as for the 1D DMRG algorithm) preset at the start of the 
calculation, size. It has thus been proven that it is possible 
to construct a multi-chain DMRG approach for 2D lattices 
which uses previously determined system and environment 
blocks {\it at all points}. 

This new 2D DMRG approach tested for the 
spin-half and spin-one square-lattice Heisenberg models
in order to predict their ground-state energies with great 
success. The best results for ground-state energy of 
the square-lattice Heisenberg model for the $N=20 \times 20$ 
lattice for the spin-half and spin-one models were found 
$E_g/N_B = -0.3321$ and $E_g/N_B = -1.1525$. 
The results for the Heisenberg model on the 
$N=6 \times 6$ and  $N=12 \times 12$ were found to be 
a reasonable agreement with both previous DMRG results for the 
spin-half square-lattice model \cite{dmrg_2D_5} with 
$M=50$ and $M=100$. Furthermore, the new DMRG results
for the $N=20 \times 20$ lattice were seen to be in reasonable
agreement with the results of the best of other approximate 
methods \cite{qmc4,series1,swt,ccm1,ccm2} determined 
in the limit $N \rightarrow \infty$. These new results, 
and those results of previous and highly successful DMRG 
calculations in 2D \cite{dmrg_2D_1,dmrg_2D_2,dmrg_2D_3,dmrg_2D_4,dmrg_2D_5},
now show that the DMRG method is quickly becoming 
a powerful tool in order to simulate and understand strongly
interacting two-dimensional quantum many-body systems. 

Future 2D DMRG calculations for the spin-half and
spin-one Heisenberg models are envisaged for even 
larger square lattices and for larger values for $M$. 
Other future calculations using the new DMRG approach 
could also be for the spin-half and spin-one Heisenberg models
on the triangular and Kagom\'e lattices. Indeed, 
extensions of the approach presented in this 
article to lattice quantum spin systems with 
frustrating next-nearest-neighbour bonds are
also possible. It is finally noted that the 2D
algorithms considered in this article may be
applied to a wide range of both fermionic and 
bosonic systems.

\vspace{0.5cm}

{\bf Acknowledgement:} I wish to gratefully acknowledge and 
to thank Ulrich Schollw\"ock for his strong 
support in writing this article and for his invaluable
help and advice in performing DMRG calculations. 

\pagebreak

\appendix

\section{Density Matrix Renormalisation Group (DMRG) Algorithm for 
 two spatial dimensions}

\begin{figure}
\epsfxsize=8cm
\centerline{\epsffile{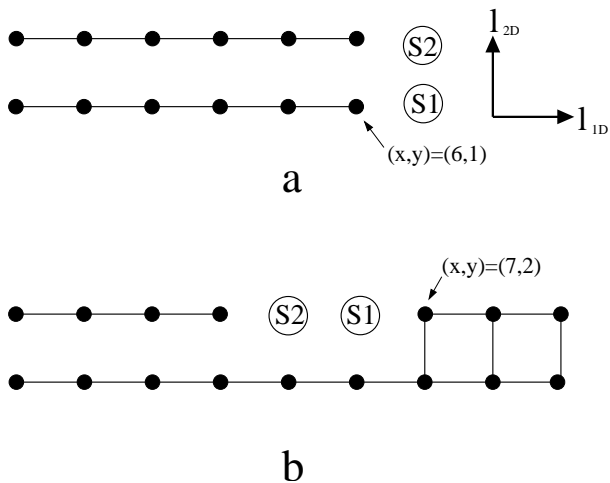}}
\vspace{0.4cm} 
\caption{The one-dimensional system block is the lower block 
 of sites and it is used to initialise the infinite-lattice 
 algorithm. The ``free'' spins $S1$ and $S2$ are indicated by the unfilled 
 circles and the environment is the upper block (chain) of sites. Diagram 
 (a) thus shows a one-dimensional system block $(6,1)$ of width $l_{1D}(=6)$ 
 and of height $1$. Figure (b) shows the system block $(7,2)$ which is
 of width $L(=9)$ and height $2$. (Note that the {\it differently
 shaped} blocks are uniquely described by the indices $(x,y)$
 shown above.)} 
\label{fig2}
\end{figure}

The method of building up the 2D lattice is now 
introduced and the first step in this algorithm is
to build up {\it one-dimensional lattices} of {\it width}
$L$ and {\it height} of one and two sites (shown in
Fig. \ref{fig2}). These blocks are used as the initial
starting point for the two-dimensional infinite-lattice
algorithm. Note that each {\it differently shaped} block 
in Fig. \ref{fig2} is uniquely described by the indices 
$(x,y)$, as shown in the figure and that this labelling
of the differently shaped blocks is crucial to this 
2D implementation of the DMRG algorithm. System blocks are 
thus uniquely denoted by the indices $x$ and $y$, where $x$ 
which refers to the right-most site for the $y$-th ``row'' 
of sites from the bottom of the system block. By comparison, 
the index $x$ represents the $x$-th left-most site for the $y$-th 
``row'' of sites from the top of the environment block in this 
case. Thus, system and environment blocks with the same indices 
refer to block shapes which are related by a transformation 
of a rotation of 180$^{\circ}$ (about some arbitrary 
central point).

The density matrix for the free spin $S1$ and the system 
block is obtained by integrating out the degrees of freedom 
of the other spin $S2$ and the environment block. The new 
Hamiltonian and spin operators are then determined in terms 
of this ``augmented block'' of the system block and its 
neighbouring spin $S1$. All of the information regarding 
these system blocks of shape ($x$,$y$) (for example, the 
number of states, the Hamiltonian, and the spin operators) 
is saved to disk for both of bottom and top blocks. This 
information is used later in order to start the full 2D 
lattice infinite-lattice algorithm. Hence, note 
that one firstly grows the {\it width} of the lattice until 
it reaches a preset size $L$ and height 1, as shown in 
Fig. 1a. One then `sweeps' from the left to the right such 
that the free spins S2 and S1 run along the top of the 
ladder-like structure shown in Fig. 1b. The quasi-2D
blocks are now of height 2 and the process is stopped 
once the spin S2 lies on the left-most site.

\begin{figure}
\epsfxsize=8cm
\centerline{\epsffile{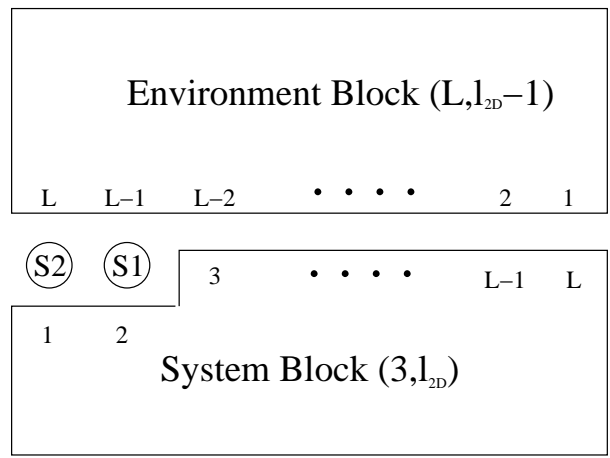}}
\vspace{0.4cm} 
\caption{The first step in the infinite-lattice
 approach. Note that the positions of the sites in the bottom-most 
 row of the environment block are reversed from left to right. ``Free''
 spins $S1$ and $S2$ are again indicated by the unfilled circles.} 
\label{fig3}
\end{figure}

In the first truly two-dimensional ``step,'' the system block 
of index $(x,y)=(1,L)$ is copied into the environment block, 
although the ordering of the sites on the bottom row of 
the environment block are {\it reversed} such 
that sites in the environment block run from $L, ~L-1, 
~\cdots,~ 1$, shown in Fig. \ref{fig3} for $l_{2D}=2$. 
The reason why one perhaps ought to do this is 
because it is hoped that it is beneficial to use a 
block of width $L$ in order to retain the maximum 
number of bonds between the system and environment block 
(namely, $L$ of them). Note that the system block 
is denoted by indices $(3,2)$ (again see Fig. \ref{fig3}).

Presumably, the symmetries of this ``reversed'' block are 
slightly different from that of the unreversed case,
and so this choice of environment block it is not ideal.
However, the goal of this approach is to use previously
defined blocks in order to quickly build up the final 
lattice in order for the finite-lattice algorithm to 
find the fixed-point of the RG process. However, note that 
that there may exist many alternate ways of performing 
this crucial ``first step'' and more is said of this below.
Now that the first two-dimensional step has been described
in detail it is possible to present the full algorithm. 
Note that we start from step 2 after defining the
initial step mentioned above for the augmented block $(2,2)$
and that the density matrix for the system block and 
spin $S1$ and the relevant Hamiltonian and spin operators 
obtained for this augmented block are stored to disk at all
points. The `steps' in the 2D algorithm is now given by:

\begin{figure}
\epsfxsize=8cm
\centerline{\epsffile{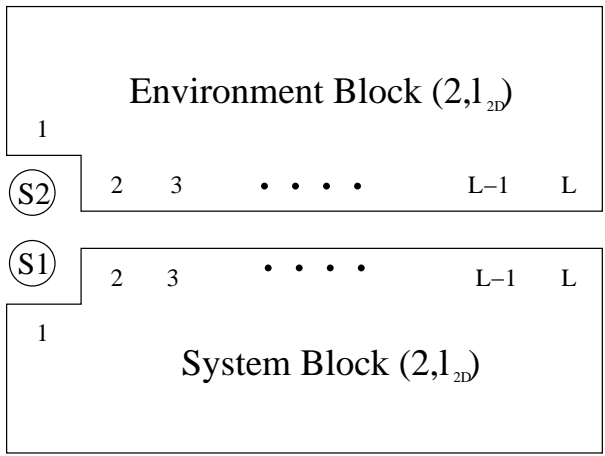}}
\vspace{0.4cm} 
\caption{The second such ``step'' in the infinite-lattice
 approach. Note new system block determined in the previous step, 
 namely (2,$l_{2D}$), is used in both the system and environment
 blocks.} 
\label{fig4}
\end{figure}

\begin{figure}
\epsfxsize=8cm
\centerline{\epsffile{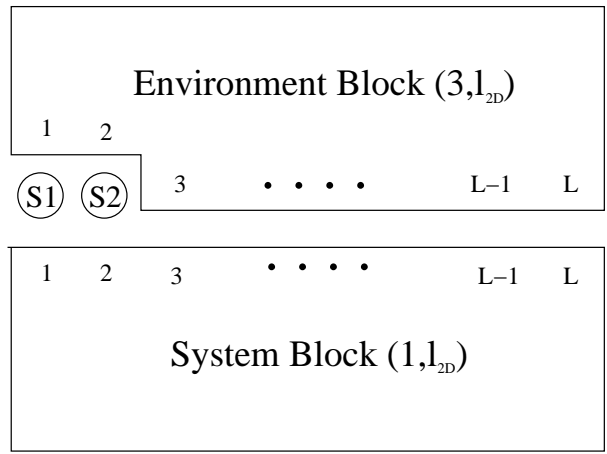}}
\vspace{0.4cm} 
\caption{The third ``step'' in the infinite-lattice
 approach.} 
\label{fig5}
\end{figure}

\begin{figure}
\epsfxsize=8cm
\centerline{\epsffile{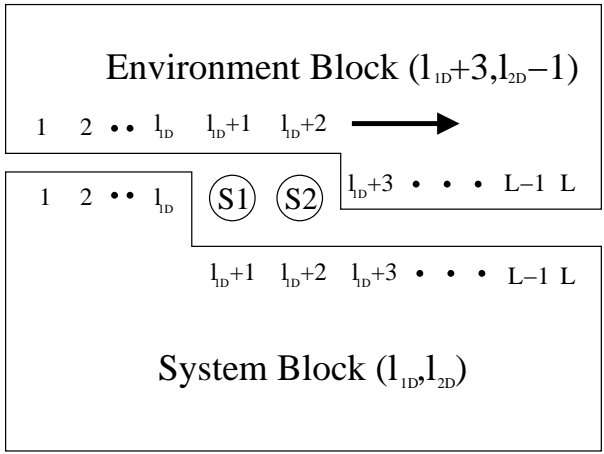}}
\vspace{0.4cm} 
\caption{The fourth and final set of ``steps'' infinite-lattice
 approach. The arrow shows the direction of the sweeping and 
 note that $l_{1D}$ runs over all indices $\{1,2,~\cdots,~L-3\}$.} 
\label{fig6}
\end{figure}

\begin{enumerate}

\item The system block of index  $(x,y)=(l_{2D}-1,L)$ is now 
copied into the environment block and the ordering of the 
sites on the bottom row of the environment block are 
reversed such that sites in the environment 
block run from $L, ~L-1, ~\cdots,~ 1$, shown in Fig. \ref{fig3}.
The two ``free'' spins are put at positions $(1,l_{2D})$ and 
$(2,l_{2D})$ and the system block is defined by the 
left-most site in the second row, namely, block 
$(x,y)=(3,l_{2D})$. The density matrix for the system 
block and the spin $S1$ is obtained and the effective
Hamiltonian and spin operators determined.  The results 
are then saved for {\it both} the top and bottom blocks 
of this shape of the augmented block, namely, 
$(x,y)=(2,l_{2D})$.

\item The next 2D ``step'' shown in Fig. \ref{fig4} 
uses system and environment blocks 
{\it both} indexed by $(2,l_{2D})$, although the sites in the
bottom row of the environment block are not reversed in
this case. Again, the density matrix for the system 
block and the spin $S1$ is obtained and the effective
Hamiltonian and spin operators determined. This information
is saved to disk for the augmented block(s) given by 
$(1,l_{2D})$.

\item In the next step the system block is referred to by the 
index $(1,l_{2D})$ and the environment block by $(3,l_{2D})$,
as is shown in Fig. \ref{fig5}. The value of $l_{2D}$ is now
incremented.  For example, after the routine which determines 
the initial 1D blocks (see Fig. \ref{fig2}) has been
completed then $l_{2D}$ is increment from a value of 2 to 3. 

\item The fourth type of ``step'' (or, more accurately stated, a set 
of steps) is in one in which one ``sweeps'' through values of 
$l_{1D}=\{1,2,~\cdots,~L-3\}$, shown in Fig. \ref{fig6} .
The density matrices are determined at each step
and the information for the system block $(x,y)=(l_{2D},l_{1D})$ 
for {\it both} the top {\and bottom} 2D blocks is saved.

\end{enumerate}

Clearly, the four types of different step described above may 
be used in an analogous way in order to sweep leftwards. (Note 
however that one starts from step 1 from now on). 
Furthermore, the final $N=L \times L$ lattice is formed 
by repeatedly sweeping right and then left, and so on. 
Thereafter the information stored to disk may be used in order 
to implement a finite-lattice algorithm in which one sweeps 
through all lattice sites repeatedly until a fixed point of 
the RG process is reached. It is furthermore 
noted that prediction of the wave function is possible for 
both the infinite-lattice {\it and} the finite-lattice algorithms, 
and that this can reduce the number of Lancz\"os iterations by up 
to an order of magnitude.

It is hoped that this method might, in principle, 
allow one to add in the degrees of freedom associated
with the spin $S1$ into new 2D blocks in a very controlled 
and efficient manner. Furthermore, it is hoped that
the use of effective system and environment blocks 
{\it at all points} might to help in simulating the 
effects of the surrounding lattice spins on these 
free spins -- which might be especially 
pronounced for larger lattices. 

The infinite-algorithm presented here is, perhaps,
one of the simplest 2D algorithms which utilises 
previously defined fully 2D blocks in order to 
build up the 2D lattice. The research presented
in this article (and that also presented in Ref.
\cite{dmrg_2D_5}) however conclusively proves that one 
may construct DMRG infinite-lattice algorithms which
utilise previously defined 2D blocks {\it at all 
points}. Note however that more elegant solutions
which also contain this property might be constructed
using similar ideas, as mentioned above. For example, 
an environment block of shape $(1,l_{2D}-2)$ might be 
used in step 1 of the 2D algorithm presented here. Another
alternative is to use a line of spins (or partial line
of spins), which would be necessary for step 1 only.
Both of these alternatives to the `step 1' used above 
would obviate the need to reverse the ordering of 
the lattice sites on the bottom on the environment 
block in this step. 

Note again that the Heisenberg Hamiltonian is given by 
Eq. (\ref{eqn1}) and that we wish to solve the Schr\"odinger 
equation, $H |\Psi\rangle = E_g |\Psi\rangle$,
where $|\Psi\rangle$ is the superblock ground-state wave function.
The Hamiltonian may also be broken into distinct parts, given by
\begin{eqnarray}
H &\equiv& H_{\rm system} +  H_{\rm system, S1}  + H_{\rm system, S2}  
\nonumber \\
&+& H_{\rm system, environment} 
 + H_{\rm environment}  \nonumber \\
&+&  H_{\rm environment, S2}  + H_{\rm environment, S2} ~~ .
\label{eqn6}
\end{eqnarray}
It is noted that (by far) the largest amount of computational time 
is spent evaluating $H_{\rm system, environment} |\Psi\rangle$
during each Lancz\"os iteration. Indeed, the number of operations 
involved in determining this scales as $L M{^4}$, where $M$
is the number of states retained at each DMRG iteration. 

All system/environment interaction terms for the $L$ distinct 
bonds between the system and environment blocks are determined 
and the gathered together before the Lancz\"os algorithm is invoked, 
and this information is then saved to local (RAM) memory. The number 
of operations involved in determining system/environment interactions
is thus reduced by a factor of $L$, although
the memory usage is commensurately increased by a factor
of $M^4$ which puts a limit on the maximum number of the
states $M$. Note that future calculations will distribute 
this memory usage up by the use of parallel processing, 
such that each processor which on a given part of the whole 
system block/environment block interactions. This not only 
reduces the memory usage of each individual machine but 
also considerably speeds up the DMRG algorithm. 

Furthermore, the data for the lower and upper two-dimensional 
blocks for indices $(x,y)$ is also saved to disk. This is 
because the memory required to store all of the information 
for the spin operators scales with $L^3 M^2$ which again very quickly 
becomes prohibitive. However, as this data is read from disk
once per DMRG iteration, this approach is not too 
inefficient. The problem of system/environment block
interactions appears to be present in {\it all} 2D implementations
of the DMRG method in 2D and so an efficient implementation 
of these terms is crucial to a good DMRG code.



\end{document}